\documentclass[fleqn,usenatbib]{mnras}

\usepackage{newtxtext,newtxmath}

\usepackage[T1]{fontenc}
\usepackage{ae,aecompl}

\usepackage{graphicx}	
\usepackage{amsmath}	
\usepackage{amssymb}	
\usepackage{multicol}
\usepackage{ulem,color}

\usepackage{caption} 

\title[Correlation]{Correlation of highly variable blazars with TeV IceCube track events}

\author[Moharana et.al.,]{R. Moharana$^1$ \thanks{E-mail:reetanjali@iitj.ac.in}, P. Majumdar$^{2}$, P. P. Basumallick$^{2}$, D. Bose$^{3}$, R. Prince$^{4}$, N. Gupta$^{4}$
\\
$^{1}$Indian Institute of Technology Jodhpur, India.\\
$^{2}$ Saha Institute of Nuclear Physics, HBNI, Kolkata, West Bengal 700064, India.\\
$^{3}$ Indian Institute of Technology Kharagpur, India.\\
$^{4}$ Raman Research Institute, C.V. Raman Avenue,
Sadashivanagar, Bangalore 560080, India.
}

\pubyear{2019}

\begin{document}
\label{firstpage}
\pagerange{\pageref{firstpage}--\pageref{lastpage}}
\maketitle

\begin{abstract}
The first ever identification of a cosmic ray accelerator as the consequence of spacial and temporal correlation of IceCube event 170922A with flaring of a blazar TXS 0506+056 motivated us to look for other flaring blazars in Fermi-LAT 3FGL catalog, which could be correlated with IceCube high energy track events. We have studied the Fermi-LAT light curves of blazars correlated with neutrino track events. 
Among the eight sources identified within 2$\sigma$ angular
 uncertainty of the IceCube track events selected in our study, we find only one source 3FGL J2255+2409 was
  in flaring state during the neutrino detection. We have  carried out a time dependent modelling of the multi-wavelength data from this blazar, and the neutrino event including leptonic energy losses and proton-proton interactions in its jet to determine whether it could be the origin of the neutrino event. Our lepto-hadronic model estimates a jet luminosity of $L_j = 3.6 \times10^{47}$ erg/sec during the neutrino phase of 3FGL/4FGL J2255+2409.

\end{abstract}


\maketitle
\flushbottom
\section{Introduction}
 
Highest energy cosmic rays are believed to be accelerated in extra-galactic astrophysical sources. High energy neutrinos are produced in or near these sources when cosmic rays interact with matter and ambient radiation. Unlike cosmic rays, these secondary neutrinos are neutral and very weakly interacting, as a result they travel through space undeflected from the direction of their sources to us. Therefore they are assumed to be smoking gun evidence for tracing origin of cosmic rays. Among different extra-galactic sources, blazars, a class of AGN with their powerful relativistic jets pointed towards us, are likely candidates for the sources of high energy cosmic rays.
  
IceCube neutrino telescope at the South Pole has detected many neutrino events of astrophysical origin over the last decade.
On 22nd September, 2017 IceCube detected neutrino track-like event (IceCube-170922A) with energy $> 290$ TeV \footnote{https://gcn.gsfc.nasa.gov/gcn3/21916.gcn3}, which was coincident in direction and time with
the $\gamma$-ray flare from a blazar TXS 0506+056.
Subsequently a multi-wavelength campaign was followed involving telescopes across the globe. Most importantly high energy gamma ray flux $< 4.5 \times 10^{-11}$
cm$^{-2}$ s$^{-1}$ above energy $90$ GeV was detected by MAGIC detector after stacking two epochs of observations one on Oct 3-4, 2017 and one on Oct 31, 2017 \citep{Ansoldi2018,IceCube:2018dnn}. 
IceCube collaboration analysed 9.5 years of data from the direction of TXS 0506+056 and found there was an excess of events above atmospheric background between September 2014 and March 2015 \citep{IceCube:2018cha}. These results indicate that blazars could be sources of astrophysical neutrinos. This correlation has been further analysed \citep{Padovani:2018acg}.
The possibility of more such correlations has also been explored \citep{Aartsen:2019gxs} recently. The two sources 1H 0323+342 and MG3 J225517+2409 showed gamma ray flares during neutrino detection with p-value of chance coincidence 8$\%$ and 4$\%$ respectively \citep{Franckowiak:2020qrq}.
  
   A series of papers has been published to explain the multi-wavelength flare and the neutrino event from TXS 0506+056 by leptonic and photo-hadronic emission \citep{Gao:2018mnu,Keivani:2018rnh,Oikonomou:2019djc,Xue:2019txw}, and also proton-proton interactions \citep{Sahakyan:2018voh, Wang:2018zln,Liu:2018utd,Banik:2019jlm,Banik:2019twt}.
Below, we discuss the results on theoretical modelling of TXS 0506+056 from a few of these papers.
A time dependent modeling of multi-wavelength flare of TXS 0506+056 by synchrotron and synchrotron self Compton emission (SSC) has been done by \citep{Gao:2018mnu} assuming the neutrino event is produced in photo-hadronic interactions. This scenario requires super-Eddington jet power to explain the neutrino event. X-ray emission constrains the photo-hadronic model of neutrino production. \citep{Keivani:2018rnh} considered synchroton and external Compton emission of relativistic electrons to explain the multi-wavelength spectrum of flare from TXS 0506+056 and radiatively subdominant hadronic emission to explain the neutrino event IceCube-170922A. In their model the ratio of luminosities in protons to electrons is large, $L_p/L_e \sim 250$ to 500.

Proton-proton interactions inside the jet, where the number of protons is determined from balancing the charge of electrons, can also explain the neutrino event IceCube-190722A \citep{Banik:2019jlm}. Motivated with this, we look for further candidate variable blazar sources with the publicly available high energy IceCube neutrino track events. Our angular correlation study resulted correlation of eight flaring blazars. Out of which we found blazar 3FGLJ2255+2409 has an additional correlation of flaring state in gamma-rays with the IceCube neutrino event.  We model the multi-wavelength spectra of 3FGLJ2255+2409 with time dependent leptonic model and proton-proton interactions in the emission region inside the jet to estimate the energy budget required to explain the emission. 
\section{IceCube Track Events \& Fermi-LAT variable Sources}

 IceCube has been detecting neutrinos of astrophysical origin since 2010. Most of these are shower-like events, which has on the average $15 ^{\circ}$ angular uncertainties. Therefore it is very difficult to identify the sources of these neutrino events. The track events, on the other hand, are generated by muon type neutrinos ($\nu_{\mu}$) in charged current (CC)
 interaction, having on an average of $1 ^{\circ}$ uncertainty in angular position. Hence for the analysis presented in this paper, we select all the track-like high energy events detected by IceCube from publicly available data . 
 
 Discovery of cosmic neutrino events by the World's largest neutrino telescope the IceCube Neutrino Observatory, are the first astrophysical neutrinos ever detected (see \cite{Aartsen:2014gkd,Aartsen:2015zva}). The six year high energy starting events (HESE) include 82 events that include atmospheric muon background ($25\pm 7.3$) and atmospheric neutrino background ($15.6^{+11.4}_{-3.9}$). 
Due to lesser angular uncertainty we have chosen the neutrino track events for our analysis. Out of the HESE 82 events 22 events are track events 

More HESE and EHE events have been reported in recent past for real time analysis {\footnote{https://gcn.gsfc.nasa.gov/amon\_hese\_events.html, https://gcn.gsfc.nasa.gov/amon\_ehe\_events.html}}, we have included the tracks of these events for our analysis, till November 2017. Apart from these events we have included the six year up going muon track events that have been reported by IceCube above deposited energy 200 TeV (\cite{Aartsen:2016xlq}). Out of the 29 such events 4 events have angular uncertainty more than 4 degrees, hence we have excluded these 4 events. 
 
 We have considered all the AGN (Active Galactic Nuclei) from 3FGL catalog \citep{2015ApJS..218...23A} having variability index more than 100 as possible sources of the IceCube neutrino events. 
A source is identified as variable with 99\% CL if its variability index is more than 71.44.
  445 out of 1749 AGN have variability index more than 100. 
 We looked into Fermi 3FGL catalogue and selected blazars which are located within 2$\sigma$ angular distance from these IceCube events. 

We find 8 sources correlate within the 2$\sigma$ angular uncertainty of the list of IceCube track events chosen. The statistical significance of this correlation is calculated using \citet{Moharana:2016mkl}. We found a pre-p-value $0.088$. The correlated AGN are listed in table \ref{tab1}.

\begin{table}
\begin{tabular}{|c||c|c|c|c|}
\hline
\textbf{Name} & \textbf{Associated}& \textbf{Var.} &\textbf{Ang.}  &  \textbf{MJD} \\
 & \textbf{source}& \textbf{index} &\textbf{dist.} &  \\
\hline
J1848+3216      & B2 1846+32A  & 273.125 & 0.12$^{\circ}$ & 56859 \\
J1040+0617   & GB6 J1040+0617 & 124.921&  0.27$^{\circ}$  & 57000\\
J2255+2409 & MG3 J225517+2409 & 103.265&  1.21$^{\circ}$  & 55355\\
J0823-2230     & PKS 0823-223  & 121.623&  1.56$^{\circ}$   & 55387\\
J1310+3222      & OP 313  & 556.442     &  1.00$^{\circ}$  & 56062 \\
J0509+0541    & TXS 0506+056 & 285.297 & 0.59$^{\circ}$  & 58018 \\
J2254+1608    & 3C 454.3 & 60733.9  &  0.93$^{\circ}$  & 57305 \\
J0112+3207   & 4C +31.03  & 955.465 & 0.94$^{\circ}$  & 57614  \\
\hline
\end{tabular}
\caption{{\label{tab1}}List of blazars correlated with the IceCube track events within angular uncertainty of $2\sigma$.}
\end{table}
\section{Data Analysis}
\subsection{Fermi-LAT data analysis}
 The Fermi-LAT is a $\gamma$-ray telescope in space
sensitive to photon energies greater than 50 MeV with a field
of view of about 2.4 sr \citep{Atwood2009}. The primary
observation mode of Fermi-LAT is survey mode in which the
LAT (Large Area Telescope) scans the entire sky every 3 hr. 
We analysed data for selected sources using Fermi Science Tools software package. 
In this analysis, we have considered photons of energy between 100 MeV and 500 GeV.

In our analysis of the $\gamma-$ ray data from 4FGL J2255, we have used the version ScienceTools-11-04-00 
of the software package, Fermi ScienceTools that is dedicated for the analysis of the LAT-data.  
Here, we use the recently released, fully reprocessed Pass8 dataset 
that provides an improved event reconstruction, a wider energy range, a better measurement 
of the (reconstructed) energies and a significantly increased effective area, especially in the 
low energy range. As the region of interest (ROI), we have chosen a region covering $10^{\circ}$ radius 
centred on the source by the use of the ``gtselect'' option of the Science Tools. By using the same 
tool, we also apply a cut of 500 GeV on the reconstructed energy of the photon events. 
The albedo contamination is avoided by rejecting the events with their zenith angles satisfying 
$\theta <  90^\circ$ , and also by selecting the good time intervals (GTIs) by using the ``gtmktime''
filter as suggested in the ScienceTools. Next,we analyse the dataset by using the "unbinned likelihood 
technique'', as implemented in the ScienceTools. In this analysis, we have used the photon events 
P8R3\_SOURCE\_V2 of the event class 128 and the photon to (e$^+$,e$^-$) pair conversion type 3, in 
which the pair conversion is supposed to take place both at the FRONT and the BACK tracker-layers of the 
LAT, so that the overall LAT instrument has a good point spread function(PSF) at low energy, 
simultaneously presenting a large effective area at high energy. While selecting the photon events 
as above, we adopt an ``instrument response function (IRF)'' suggested in version P8R2\_SOURCE\_V6 (check this) 
of the appropriate manual of the ScienceTools. The variability of the source was investigated 
by producing lightcurves with different time bins (daily, weekly and monthly) where upper limits 
to the flux were computed for cases where TS $<$ 4. In addition we performed the 
spectral analysis of the source in the energy range 100 MeV to 500 GeV and computed the spectral parameters
of the source. In this case also, upper limits to the flux were obtained for each spectral bin where the 
TS was $<$ 4.  

\begin{figure}
\includegraphics[width=1\linewidth]{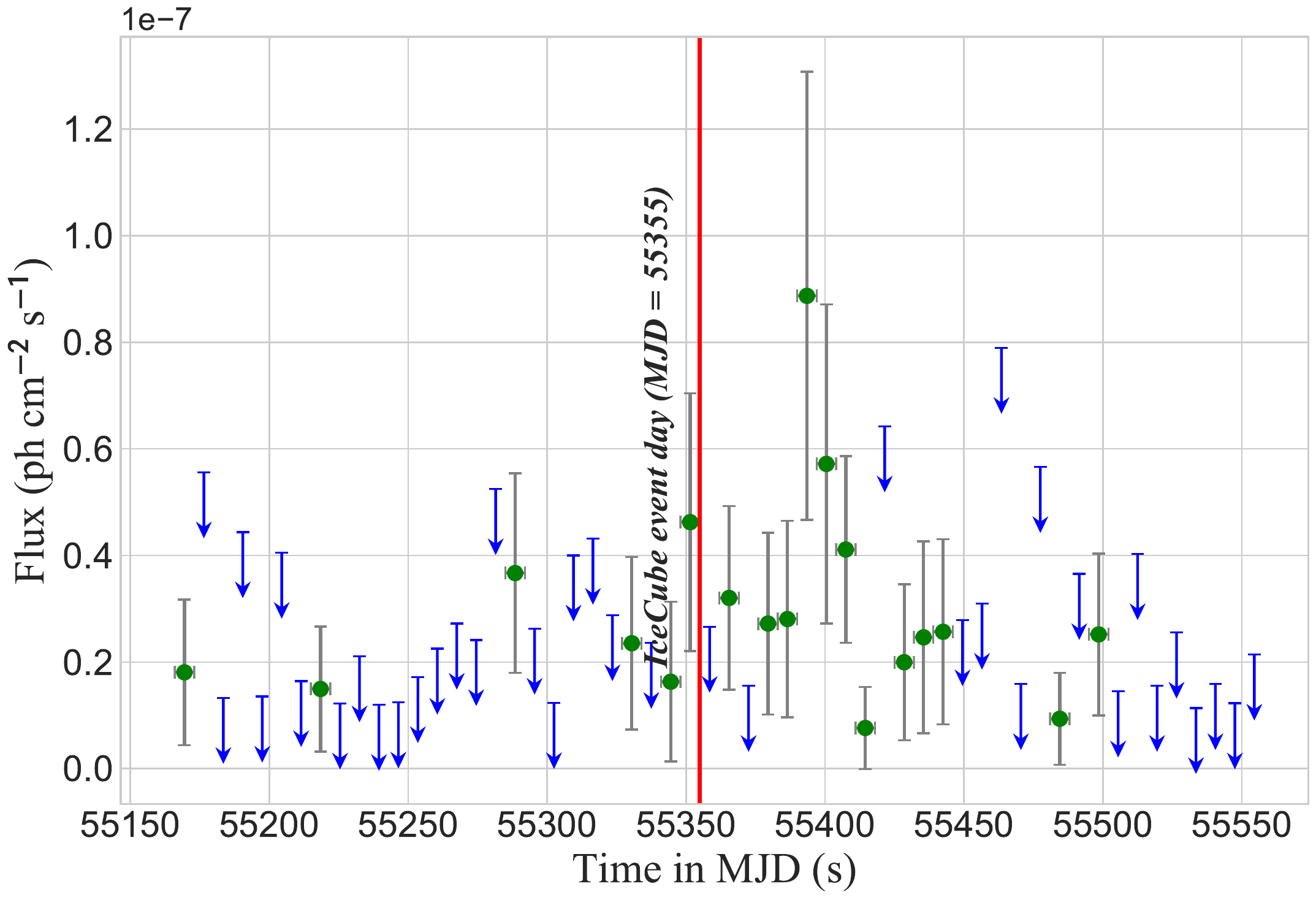}
\caption{The weekly light curve (ph/cm$^2$/s) for 4FGL J2255+2409 from 100 MeV to 500 GeV and red line shows the date of IceCube neutrino event.}
\label{lc}
\end{figure}

 We analysed all the 7 correlated sources \footnote{we have excluded the TXS0506+056 source from further analysis, as a lot of work has already been done on the source.} to look the light curve for two months within the IceCube neutrino track event and found no variability around the event except for 3FGL/4FGL J2255+2409 as shown in figure \ref{lc}.
 
 4FGLJ2255+2409 (associated source MG3 J225517+2409) is within the 2$\sigma $ angular resolution of muon track event having most probable energy 339 TeV leading to most probable $\nu$-event energy 442 TeV. This event was observed during IC79 configuration with signal probability 0.86  {\footnote{https://icecube.wisc.edu/icecube/static/science/HE\_NuMu\_data-table.pdf}} \citep{Aartsen:2016xlq}. 

There has been observational data of MG3 J225517+2409 at radio frequency 1.4 GHz from NRAO VLA Sky Survey in 1998 \citep{Condon:1998iy} and the X-ray data taken by ROSAT \citep{Massaro:2008ye}. For the spectral energy distribution analysis we collected Fermi-LAT $\gamma-$ray data for three phases, a) pre $\nu$-phase from November 2009 to April 2010 for six months b) $\nu$-phase, May to July 2010, two months and c) post $\nu$-phase, August 2010 to January 2011, another six months.

\subsection{Swift-XRT/UVOT Analysis}
We analysed Swift-XRT/UVOT data for the source 3FGL J2255+2409 during the period overlapping with IceCube event. Details on the observations are presented in Table 2.

For this we retained cleaned event files using the task "xrtpipeline"
version 0.13.2. Latest calibration files (CALDB version 20160609) and standard screening criteria were used for re-processing the raw data. 
Circular regions of radius 20 arcsec centered at the source and slightly away from
the source were chosen for the source and the background respectively. 
Source was very faint in X-ray thus not visible. Therefore we could not obtain spectra for this source.
Ultraviolet/Optical Telescope used all six filters: U, V, B, W1, M2, and W2
to observe this source.
The source image was extracted from a region of 5 arcsec
centered at the source. The background region was chosen with a
radius of 10 arcsec away from the source. The "uvot source" task has been 
used to extract the source magnitudes and fluxes. 
Magnitudes are corrected for
galactic extinction \citep{Schlafly_2011} and converted
into flux using the zero-points \citep{Breeveld_2011} and
conversion factors \citep{Larionov:2016fxo}.

\begin{table}
\begin{tabular}{|c|c|c|c|}
\hline
\textbf{Obs} & \textbf{Exp. Time} & \textbf{Exp. Time} &  \textbf{MJD} \\
\textbf{ID} & \textbf{(XRT)} & \textbf{(UVOT)} &  \\
\hline
00041537001    & 1.1 ks  & 1.0 ks & 55463.9 \\
55463.9      & 0.5 ks  & 0.5 ks & 55490.6 \\
\hline
\end{tabular}
\caption{Swift-XRT/UVOT analysis for 3FGLJ2255+2409.}
\end{table}
 
The spectral point of the neutrino event is calculated using the exposure given in \citep{Aartsen:2016xlq}. The event was detected during the partial configuration IC79 corresponding neutrino energy 442 TeV at RA =$344.93^{+3.39}_{-2.90}$, Dec = $23.58^{+0.91}_{-1.18}$ with $90\%$ C.L. 
\subsection{Lepto-Hadronic Modelling of SEDs}
The emission region is assumed to be a spherical blob of radius $R$ moving with Doppler factor $\delta$ along the axis of the jet of the source.
The injected spectrum of relativistic electrons at energy $E_e'$ in the jet frame/comoving frame is, $\frac{dN_e}{dE_e'} \propto E_e'^{-\alpha}
$. The synchrotron and synchrotron self Compton (SSC) photon spectra from these electrons are calculated using the publicly available GAMERA code {\footnote{http://libgamera.github.io/GAMERA/}}. This code solves the time depended transport equation and gives the radiation spectrum \citep{Atoyan:1997eu}. 
Similarly, for the injected protons the spectrum at energy $E_p'$ is, $\frac{dN_p}{dE_p'} \propto E_p'^{-\alpha}$, which is used in GAMERA code to calculate the resulting secondaries (electrons, positrons and photons) produced in $pp$ interactions, where the targets protons are the cold protons in the jet.
We have calculated the secondary high energy photon, electron and positron spectrum from $pp$ interactions following \citet{Kelner:2006tc}. The average neutrino flux is 2/3 of the gamma-ray flux produced in $pp$ interactions.  We have taken the cold to hot proton density ratio as $\epsilon_{p}=900$. The resulting high energy photons are attenuated due to pair production. We have calculated the opacity of the high energy photons and the resulting secondary electron, positron emissivity following the formalism discussed in \cite{Banik:2019jlm,Aharonian1983}. Using the spectrum of secondary electrons and positrons as the injection spectrum, we have calculated the cascade of photons using GAMERA. We have also included the photons generated by the radiative losses of the electrons and positrons produced from decay of charged pions. 
As this blazar has a tentative redshift of 1.37, we have corrected the gamma ray flux for attenuation by EBL following the model by Franceschini-Rodighiero-Vaccari (FRV) \citep{Franceschini:2008tp}\footnote{http://www.astro.unipd.it/background/}.

\begin{figure}
\includegraphics[width=0.47\textwidth]{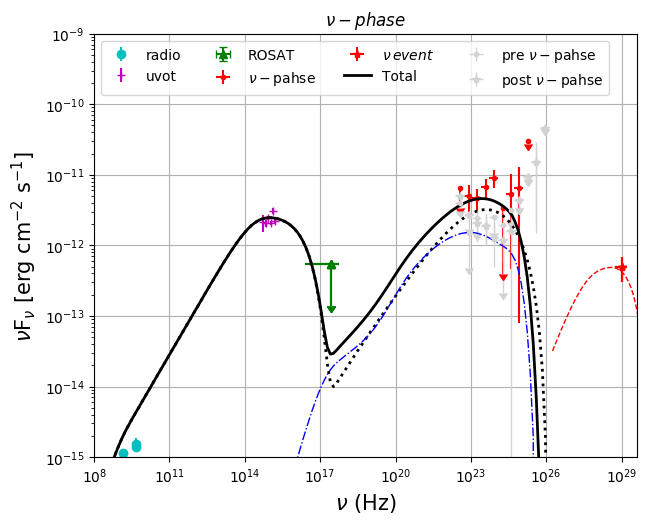}
\caption{SED modelling of the $\nu$-phase, between time period MJD55317 to MJD55407 of 3FGLJ2255+2409. The synchrotron and SSC modeling of the injected electron is shown with dotted line (black). The dashed (red) line shows the neutrino flux from pp interaction, while dot-dashed (blue) line represented the cascade photons from pp interaction. The solid line (black) shows the total flux including the EBL correction. }
\label{phase}
\end{figure}

\begin{figure}
\includegraphics[width=1.05\linewidth]{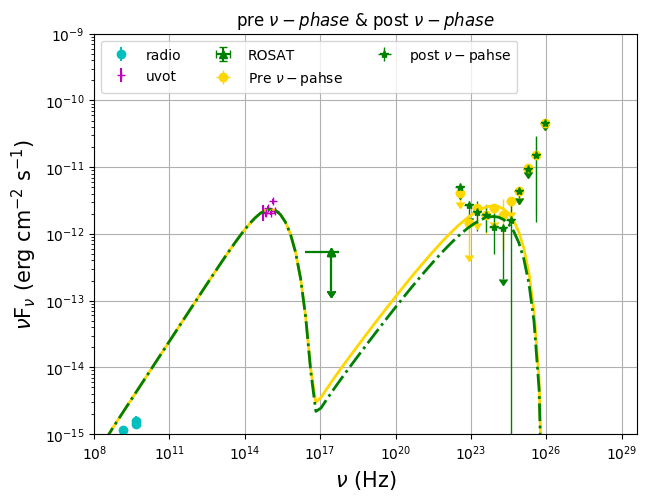}
\caption{SED modelling of the pre $\nu$-phase, within time period MJD55136 to MJD54951 with solid line (yellow) and the post $\nu$-phase, within time period MJD55409 to MJD55592 with dot-dashed line (green) of 3FGLJ2255+2409. }
\label{prepost}
\end{figure}

\begin{table}
\begin{tabular}{|c|c|c|c|}
\hline
Parameters & \textbf{pre $\nu$-phase}  & $\nu$\textbf{-phase} & \textbf{post $\nu$-phase} \\
\hline
 \textbf{$\gamma_{e,min}$} & $6.5$ & - & -\\
 \textbf{$\gamma_{e,max}$} &  $7.2 \times 10^4$ & - & - \\
 \hline
\textbf{B [G]} & $0.04$  & $0.18$  & $0.04$  \\
\textbf{$\alpha_e$} & $1.85$ & $1.78$ &  $1.85$ \\
\textbf{Radius[cm]} & $3.2 \times 10^{17}$ & $5 \times 10^{16}$ & $2.5 \times 10^{17}$ \\
\textbf{$\delta$} &  $21$ & $26.5$ & $21$ \\
\textbf{$L_{e}$ [erg/sec]}& $1.12 \times 10^{42}$& $9 \times10^{41}$ &$1.12 \times 10^{42}$\\
\hline
\textbf{Hadrons}& -& -&\\
\hline
\textbf{$\gamma_{p,min}$} & $10$ & - & -\\
 \textbf{$\gamma_{p,max}$} &  $1000$ & - & - \\
 \textbf{$\alpha_p$} & - & $1.78$ &  - \\
\textbf{$L_{p}^{hot}$ [erg/sec]} & - & $7.9 \times 10^{43}$& -\\
\textbf{$n_H^{cold}$ [cm$^{-3}$]} & - & $1466.28$& -\\
\textbf{$N_{\nu,tot}$ }& - & $1.012$& -\\
\hline
\hline
\textbf{$L_{j}$ [erg/sec]}& $8.03 \times 10^{44}$ & \boldmath$ 3.6 \times10^{47}$ & $8.4 \times 10^{44}$ \\
\hline
\end{tabular}
\caption{SED modeling parameters}
\label{mod}
\end{table}

\section{Discussion}

We studied the angular correlation of variable AGN sources from 3FGL catalogue with the IceCube TeV-PeV track events. The result showed 8 candidate sources within 2$\sigma$ angular uncertainties, including TXS 0506+056. Out of the 7 candidate sources, we found a spatial as well temporal correlation of a flare from 3FGL J2255+2409 with a 340 TeV muon track event in IceCube 79 string configuration. 

We have modeled 3FGL J2255+2409 for all the three phases, pre $\nu$-phase, post $\nu$-phase with time dependent leptonic model and the $\nu$-phase with time dependent leptonic as well time independent hardronic model using proton-proton (PP) interaction. 

The jet luminosity for the pre and post $\nu$-phase are $L_j = 8.03 \times 10^{44} $ and $8.4 \times 10^{44}$ with a blob radius $R'= 3.2 \times 10^{17}$.
The total jet luminosity during $\nu$-phase $L_j = 3.6 \times10^{47}$ erg/sec for our lepto-hadronic model is comparable to the Eddington luminosity of 3FGLJ2255+2409 $L_{Edd} = 1.2 \times 10^{47}$ erg/sec for a black hole mass of $10^9$ M$_{\odot}$. We found the neutrino event number as 1.012 for the IC79 configuration from our lepto-hadronic model in the $\nu$-phase. Our analysis shows that it does not require super-Eddington luminosity or complicated geometry to explain 3FGLJ2255+2409/MG3 J225517+2409 as the origin of the neutrino event of energy 442 TeV. To the best of our knowledge this is the first attempt to model 3FG/4FGL J2255+2409 as a source of astrophysical neutrinos.

\bibliographystyle{mnras}

\bibliography{sample.bib}

\end{document}